\documentclass[twocolumn,pra,amsmath,amssymb,aps]{revtex4}
\usepackage{dcolumn}
\usepackage{bm}
\usepackage{graphicx}
\usepackage{mathrsfs}
\usepackage{amsmath}

\begin{document}
\def\e{\mathcal{E}}

\title{Coherence assisted resonance with sub-lifetime-limited linewidth}

\author{Lei Feng}
\affiliation{Department of Physics, State Key Laboratory of Surface Physics and Key Laboratory of Micro and Nano Photonic Structures (Ministry of Education), Fudan University, Shanghai 200433, China}

\author{Pengxiong Li}
\affiliation{Department of Physics, State Key Laboratory of Surface Physics and Key Laboratory of Micro and Nano Photonic Structures (Ministry of Education), Fudan University, Shanghai 200433, China}

\author{Liang Jiang}
\affiliation{Institute for Quantum Information, California Institute of Technology, Pasadena, CA 91125, USA}
\affiliation{Department of Applied Physics, Yale University, New Haven, CT 06511, USA}

\author{Jianming Wen}
\affiliation{Institute for Quantum Information Science, University of Calgary, Calgary, Alberta T2N 1N4, Canada}

\author{Yanhong Xiao}
\affiliation{Department of Physics, State Key Laboratory of Surface Physics and Key Laboratory of Micro and Nano Photonic Structures (Ministry of Education), Fudan University, Shanghai 200433, China}

\date{\today}

\begin{abstract}

  We demonstrate a novel approach to obtain resonance linewidth below that limited by coherence lifetime.
  Cross correlation between induced intensity modulation of two lasers coupling the target
  resonance exhibits a narrow spectrum. 1/30 of the lifetime-limited width was achieved in a proof-of-principle experiment where
  two ground states are the target resonance levels. Attainable linewidth is only limited by laser shot noise in principle.
  Experimental results agree with an intuitive analytical model and numerical calculations qualitatively.
  This technique can be easily implemented and should be applicable to many atomic, molecular and solid state
  spin systems for spectroscopy, metrology and resonance based sensing and imaging.

\end{abstract}


%

\maketitle


Achieving narrow resonance is of long lasting interest to both fundamental and applied sciences such as spectroscopy, precision measurement, metrology, and sensing. Resonance lines are often broadened, for example, by Doppler effects, inhomogeneous local fields, power broadening, etc. Techniques to reduce these effects include laser cooling and trapping of atoms, rephasing via pulse sequences and Ramsey type pump-probe method~\cite{spectroscopy}. Nonetheless, resonance linewidth is ultimately limited by the lifetimes of involved atomic states, which determine the natural linewidth or more broadly the lifetime-limited linewidth. How to go beyond this limit has been a long-standing important challenge. Existing efforts can be grouped into two categories. In the first category, subnatural optical resonance is enabled by a much narrower resonance coupled to the target resonance \cite{Gawlik,Lam,Gauthier}. There, the attainable linewidth cannot beat the natural width of the narrower resonance. In the second category, a time resolved pump-probe method in the same spirit of the Ramsey spectroscopy is used to detect the longer-lived subset of an atomic ensemble \cite{Phillips,Knight,Shimizu,Scully,Albrecht}.

We here report a new method which uses continuous-wave (CW) lasers to obtain linewidth far below the coherence lifetime limit. Our technique utilizes coherence between the two lower states in a three-level system exhibiting electromagnetically induced transparency (EIT)~\cite{EITreview} or coherent population trapping (CPT)~\cite{CPTreview}. The resonance of interest is between two lower states and the attainable linewidth is much narrower than the width set by the lifetime of the ground state coherence. The narrowing factor is only determined by the laser frequency modulation parameters, and is limited by technical noise and eventually laser shot noise. In a proof-of-principle experiment, we have observed 1/30 of the lifetime-limited width by employing frequency modulated CW lasers. Since this technique is easy to implement and only requires a Raman/CPT process which exists in many systems such as atoms, molecules, and solid state spins, we believe it will have broad applications.

\begin{figure}[t]
\includegraphics[clip,width=0.6\linewidth]{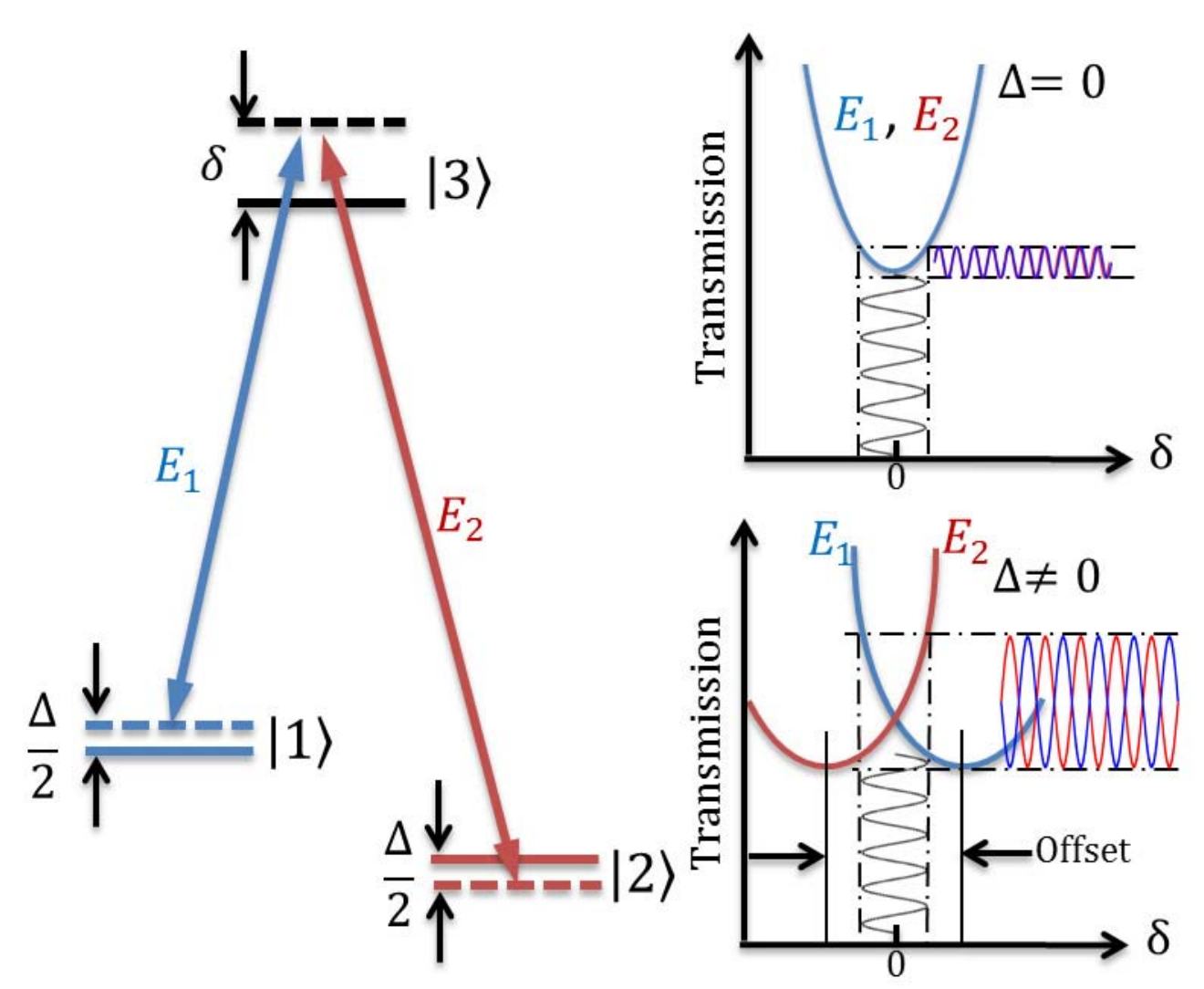}
\caption{(Color online) Principle of the sub-lifetime-limited resonance, shown between $|1\rangle$ and $|2\rangle$ in a three-level $\Lambda$ configuration. $\delta$ is the averaged one-photon detuning, and $\Delta$ is the two-photon detuning, generated by shifting the energy levels with a magnetic field via Zeeman shift (as shown) or by varying the frequency difference of $E_1$ and $E_2$. \it{(See text for details).}} \label{illustration.fig}
\end{figure}

The principle of this technique is illustrated in Fig.~\ref{illustration.fig}. Two metastable ground states $|1\rangle$ and $|2\rangle$ are resonantly coupled to an auxiliary fast-decaying level $|3\rangle$ by two optical fields $E_1$ and $E_2$ under same frequency modulation (FM). As the two-photon detuning $\Delta$ is changed from zero to nonzero, cross correlation between converted intensity (amplitude) modulations (AM) \cite{Camparo} in $E_1$ and $E_2$ exhibits a sharp transition from correlation to anti-correlation. The physics can be intuitively understood as follows. We examine the individual transmissions of $E_1$ and $E_2$ as their frequencies are slowly swept together. For $\Delta=0$, the two transmission spectra overlap and their converted AM are correlated. For $\Delta\neq0$, the two transmission spectra shift to opposite directions due to the sharp dispersion of EIT. This effectively induces an offset between the two spectra minima much greater than $\Delta$ (expected offset if there were no coherence between $|1\rangle$ and $|2\rangle$), creates a considerable range around $\delta=0$ where the two spectra possess opposite slopes, and leads to anti-correlations. Therefore, it is precisely the ground-state coherence that enables the sharp transition from correlation to anti-correlation. This picture also suggests that the transition linewidth decreases with reduced FM, as shall be verified in our theory and experiment presented below. We note that such a transition was previously addressed in EIT with noisy diode lasers \cite{CorrelationStudy,Sautenkov,XiaoPRA} but without identification of the sub-lifetime-limited width transition characteristic.

The dynamics of the generic three-level $\Lambda$ system is described by the Hamiltonian:
\begin{equation}
H=\Omega_1|3\rangle\langle1|+\Omega_2|3\rangle\langle2|-\delta|3\rangle\langle3|+\frac{\Delta}{2}(|1\rangle\langle1|-|2\rangle\langle2|)+H.c.,
\end{equation}
where $\Omega_{1,2}$ are Rabi frequencies of $E_1$ and $E_2$, and $H.c.$ represents the Hermitian conjugate. Here, $\Omega_{1,2}=\Omega_{r_{1,2}}e^{i\lambda\sin{\nu t}}$ with $\nu$ and $\lambda$ respectively the modulation frequency and modulation depth. The decay rate of $|3\rangle$ is $\Gamma$ ($\gg\Delta$), and the transverse and longitudinal relaxation rates between ground states are $\gamma_2$ and $\gamma_1$. The zero time lag intensity cross correlation between $E_1$ and $E_2$ is defined by $g^{(2)}(0)\equiv\langle\delta I_1(t)\delta{I}_2(t)\rangle/\sqrt{\langle(\delta{I}_1)^2\rangle\langle(\delta{I}_2)^2}\rangle$, where $\delta I_{1,2}$ denote intensity fluctuations of the two transmitted fields after atoms. The positivity (negativity) of $g^{(2)}(0)$ means that two output fields are correlated (anti-correlated). For an optically thin medium, $g^{(2)}(0)$ can be expressed as $g^{(2)}(0)=\langle\delta\rho^i_{31}\delta\rho^i_{32})\rangle/\sqrt{\langle(\delta\rho^i_{31})^2\rangle\langle(\delta\rho^i_{32})^2}\rangle$ \cite{Sautenkov}, where $\langle\cdot\rangle$ is a time average, and $\delta\rho^i_{31}$ and $\delta\rho^i_{32}$ are the imaginary
parts of fluctuations away from the steady states of the atomic density matrix elements.

To gain physical insights, we pursued analytical solutions in both adiabatic and nonadiabatic regimes by assuming $\Omega_{r1,r2}=\Omega$ and equal decay rates from $|3\rangle$ to $|1\rangle$ and $|2\rangle$. Without these assumptions, our numerical simulations also confirmed the subnatural linewidth characteristic. When the modulation frequency $\nu$ is much lower than the optical pumping rate and the modulation amplitude $\lambda\nu$ is much smaller than $\Gamma$, atoms adiabatically follow the steady-state solution:
\begin{equation}\label{adiabatic}
\begin{aligned}
\rho^i_{13}=\frac{-\Omega}{\Gamma(1+\delta^2_0)}\left[\left(\frac{1}{2}+\rho^r_{12}\right)+\delta_0\left(1-\frac{2\Gamma_p}{\gamma_1+2\Gamma_p}\right)
\rho^i_{12}\right],\\
\rho^i_{23}=\frac{-\Omega}{\Gamma(1+\delta^2_0)}\left[\left(\frac{1}{2}+\rho^r_{12}\right)-\delta_0\left(1-\frac{2\Gamma_p}{\gamma_1+2\Gamma_p}\right)
\rho^i_{12}\right],
\end{aligned}
\end{equation}
where $\delta_0=\delta/(\Gamma/2)=\lambda\nu\cos{\nu t}/(\Gamma/2)$ is the normalized one-photon detuning, $\Gamma_p=\frac{\Omega^2}{2\Gamma(1+\delta^2_0)}$ is the reduced optical pumping rate, $\rho^i_{12}$ and $\rho^r_{12}$ are the imaginary and real parts of the ground-state coherence $\rho_{12}=\frac{-\Gamma_p}{\gamma_2+2\Gamma_p+i\Delta}$. The physics becomes now clear. As seen from Eq. (2), the first (second) two terms in the square brackets correspond to the correlated (anti-correlated) intensity. A static offset between the minima of the transmission spectra of $E_1$ and $E_2$ arises from the anti-correlation terms.
For $\delta_0\ll1$, by Taylor expansion and keeping the first and second orders of $\delta_0$ in Eq. (2), we find that the correlation
(anti-correlation) terms only contain even (odd) orders of $\delta_0$. This is because the two transmission spectra have quadratic and parallel FM-AM slopes for $\Delta=0$, and have linear but opposite slopes for $\Delta\not=0$. Due to the $\rho^i_{12}$ factor, the anti-correlation terms are zero at $\Delta=0$ and then grow rapidly with increasing $|\Delta|$. When $|\Delta|=\Gamma_{g2}$, anti-correlation and correlation terms cancel and $g^{(2)}(0)=0$. Here $\Gamma_{g2}$ is the half width at half maximum (HWHM) of $g^{(2)}(0)$ which takes a simple form when $\gamma_1=\gamma_2$ and $\gamma_2\ll2\Gamma_p$:
\begin{equation}\label{width1}
\begin{aligned}
\Gamma_{g2}=\frac{\gamma_2}{\Gamma/(\lambda\nu)}\left(1+\frac{\gamma_2}{2\Gamma^0_p}\right),
\end{aligned}
\end{equation}
with $\Gamma^0_p=\frac{\Omega^2}{2\Gamma}$. Equation (3) clearly shows that $g^{(2)}(0)$ has subnatural linewidth which increases with increasing modulation depth and frequency. 

In the nonadiabatic regime, modulation frequency can be much larger than the optical pumping rate and even $\Gamma$. In this case, ground-state coherence cannot follow the laser FM, but the slopes of transmission spectra of $E_1$ and $E_2$ around $\delta_0=0$ remain parallel for $\Delta=0$ and opposite for $\Delta\not=0$. We can obtain the first- and second-order responses of the atomic coherence using perturbation theory when the modulation depth $\lambda$ is small:
\begin{equation}\label{nonadiabatic}
\begin{aligned}
 \rho^{i(1)}_{13}=-\rho^{i(1)}_{23}=-\frac{2\Omega J_1(\lambda)}{\sqrt{(\Gamma/2)^2+\nu^2}}\rho^{i(0)}_{12},\\
 \rho^{i(2)}_{13}=\rho^{i(2)}_{23}=\frac{2\Omega J_2(\lambda)}{\sqrt{(\Gamma/2)^2+(2\nu)^2}}\left(\frac{1}{2}+\rho^{r(0)}_{12}\right),
\end{aligned}
\end{equation}
where $\rho^{(n)}_{ij}$ are the amplitudes of the nth-order responses. Similar to the adiabatic case, even (odd)-order terms correspond to correlation (anti-correlation). The HWHM of $g^{(2)}(0)$ can be derived as:
\begin{equation}\label{width2}
\begin{aligned}
\Gamma_{g2}=\frac{\gamma_2}{C}\left(1+\frac{\gamma_2}{2\Gamma^0_p}\right),
\end{aligned}
\end{equation}
where
$C=\frac{J_1(\lambda)}{J_2(\lambda)}\sqrt{\frac{(2\nu)^2+(\Gamma/2)^2}{(\nu)^2+(\Gamma/2)^2}}$, and $J_l$ is the Bessel functions of the first kind. The linewidth is subnatural provided that $C>1$, which always holds for relatively small modulation depth.

\begin{figure}[t]
\includegraphics[width=0.6\linewidth]{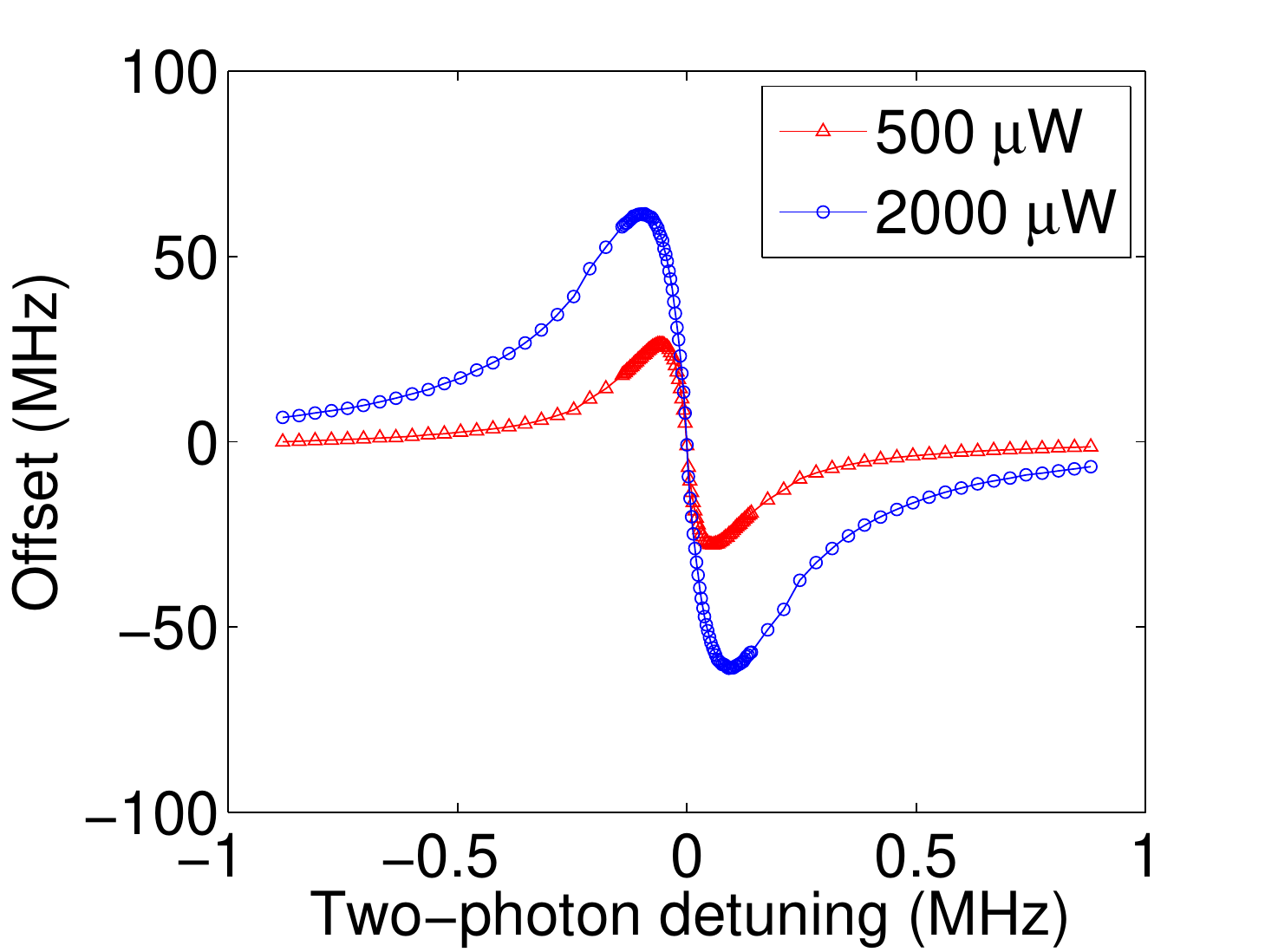}
\caption{(Color online) Measured offset between transmission minima of $E_1$ and $E_2$ vs. $\Delta$ for two input laser powers. Lines are to guide the eye.} \label{offset.fig}
\end{figure}

We performed a proof-of-principle experiment in $^{87}$Rb vapor under the CPT configuration, with ground states being Zeeman sublevels of $5^2S_{1/2}, F=2$ and excited states $5^2P_{1/2}, F=1$. The two photon detuning $\Delta$ is produced by the magnetic field. The linewidth of the CPT resonance is equal to the ground state decoherence rate plus power broadening. We used a narrow-linewidth ($<1$ MHz) diode laser, and modulated its frequency via either piezoelectric transducer (PZT) or current. Within a three-layer magnetic shield, the vapor cell was placed inside a solenoid which provided a homogeneous magnetic field. The cell temperature was maintained at 52$^{\circ}$C. The laser beam was linearly polarized before entering the cell. The left and right circularly polarized components played the role of $E_1$ and $E_2$ which were separately detected after the cell by amplified photodetectors. The AC signals were recorded by an oscilloscope and the cross correlation of intensity fluctuations was computed offline.

We first measured the offset between the minima of transmission spectra for $E_1$ and $E_2$. For a fixed two-photon detuning $\Delta$, individual transmissions were recorded as laser frequency swept slowly. As expected, the two spectra overlap at $\Delta=0$ and offset to opposite directions for nonzero $\Delta$ (see Fig.~\ref{illustration.fig}). The offset value was then retrieved by fitting the spectra with a double Gaussian profile to account for the other excited state. The offset v.s. two-photon detuning is plotted in Fig.~\ref{offset.fig}. The dispersive shape is in agreement with predictions of Eq.~(\ref{adiabatic}). Also, the offset increases with input laser power and then saturates. For 2 mW input power, a maximal offset of 60 MHz was obtained at $\Delta=0.1$ MHz, corresponding to an effective amplification factor of 600.

\begin{figure}[t]
\includegraphics[clip,width=0.8\linewidth]{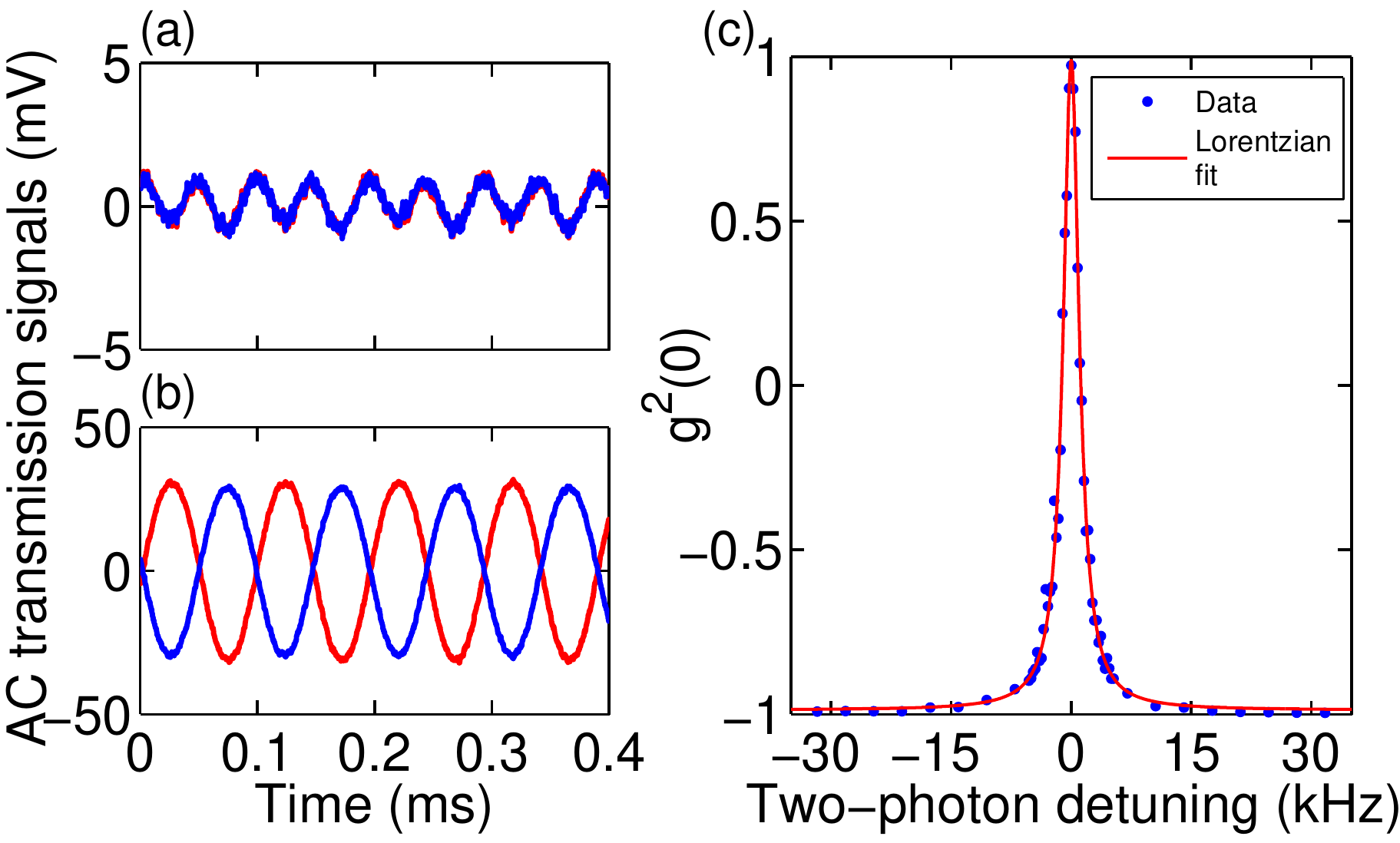}
\caption{(Color online) AC transmitted signals of the two CPT fields (blue and red) for (a) $\Delta=0$ and (b) $\Delta=15$ kHz. (c) An example of $g^{(2)}(0)$ spectrum with FWHM 2.4 kHz, about 1/30 of the transit width $\gamma_2/\pi=75$ kHz. The input laser power was 300 $\mu$W.} \label{response1.fig}
\end{figure}
Next, we looked at the $g^{(2)}(0)$ v.s. $\Delta$ spectrum in the adiabatic regime. The laser frequency was modulated by alternating the PZT voltage at $10.3$ kHz, with the FM amplitude within tens of MHz. For $\Delta=0$, we observed correlated intensity fluctuations which only have the second harmonic of the modulation frequency [Fig.~\ref{response1.fig}(a)]. At a small nonzero $\Delta$, the correlation can become completely negative and only the first harmonic appears [Fig.~\ref{response1.fig}(b)]. The amplitude of the converted AM is small at zero $\Delta$ and much larger at nonzero $\Delta$ because the FM-AM slopes are quadratic at $\Delta=0$ and linear as $\Delta\not=0$. Figure~\ref{response1.fig}(c) gives an example of $g^{(2)}(0)$ spectrum with a linewidth of 2.4 kHz. This is about 1/30 of the lifetime-limited width of 75 kHz, obtained by extrapolating the zero-power CPT width from the measured width v.s. power curve and consistent with the transit broadening estimated from the $1/e^2$ laser beam diameter of 2.2 mm.
\begin{figure}[t]
\includegraphics[width=0.8\linewidth]{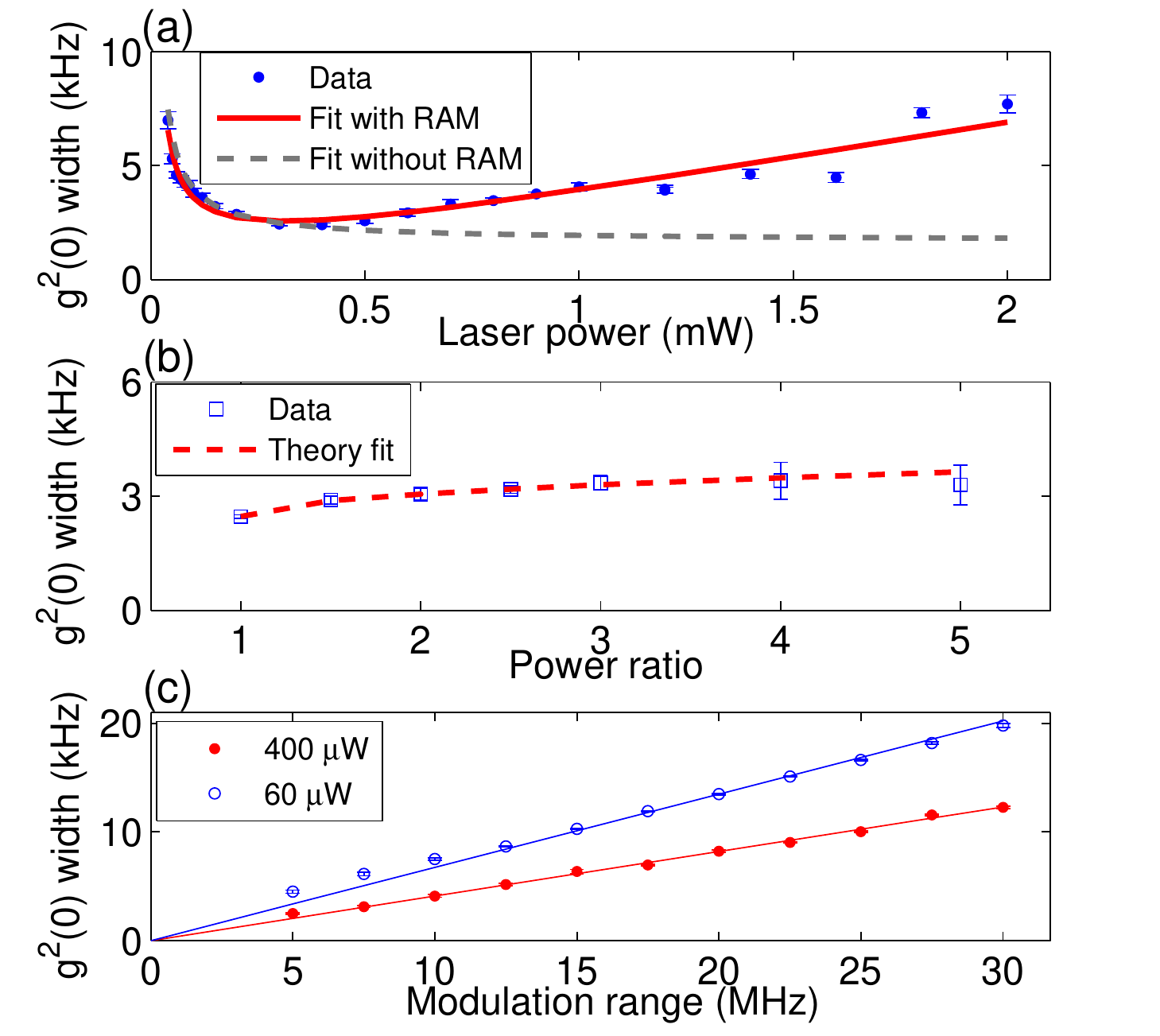}
\caption{(Color online) Dependence of $g^{(2)}(0)$ FWHM on (a) input laser power, on (b) the power ratio of the two CPT beams for laser power of 300 $\mu$W, and on (c) the modulation range for two different laser powers. Full modulation range $2\lambda\nu$ is 5 MHz for (a) and (b). Lines in (c) are linear fits. In (a) and (b), an overall scaling factor was applied to the linewidth in the fitting to account for broadening from optical depth and other noises. The transit width is 75 kHz.} \label{laserpower.fig}
\end{figure}

The dependence of the $g^{(2)}(0)$ linewidth upon the input laser power and modulation depth was studied in the adiabatic regime. As shown in Fig.~\ref{laserpower.fig}(a), the linewidth is larger at lower laser power, consistent with theoretical prediction (dashed line). This is because the
offset is small for low power and complete anti-correlation cannot be reached for the chosen range of laser FM, resulting in larger linewidth. Such an
abnormal trend is in contrast with power broadening in traditional spectroscopy. The observed width increase at higher laser power is due to influences
of the residual amplitude modulation (RAM) in the laser. Taking into account of this RAM, our simulation reproduces the experimental observation [solid line in Fig.~\ref{laserpower.fig}(a)]. The signature of the RAM's influence is the appearance of the first harmonic in the correlation signal at $\Delta=0$, as barely seen in Fig.~\ref{response1.fig}(a).
To show the general applicability of this technique, we varied the power ratio of $E_1$ and $E_2$ by placing a quarter-wave plate before they entered the cell. We found that the $g^{(2)}(0)$ width has only a small increase as the field intensities become more imbalanced,
consistent with the numerical simulation [see dashed line in Fig.~\ref{laserpower.fig}(b)]. This weak dependence on the ratio of two laser intensities
allows to optimize the ratio against other detrimental effects such as light shifts in precision measuremnts.
To check the dependence of the $g^{(2)}(0)$ linewidth on modulation depth, we tuned the PZT modulation voltage and observed the linear dependence of the width (Fig.~\ref{laserpower.fig}(c)), which agrees with the trend seen from Eq.~(\ref{width1}). The physics is that a larger range of frequency variation ``sees" more parallel slopes and thus reduces anti-correlation. For higher laser power, since the offset is larger (see Fig.~\ref{offset.fig}), $g^{(2)}(0)$ width is more ``immune" to modulation depth increase.

\begin{figure}[t]
\includegraphics[width=0.6\linewidth]{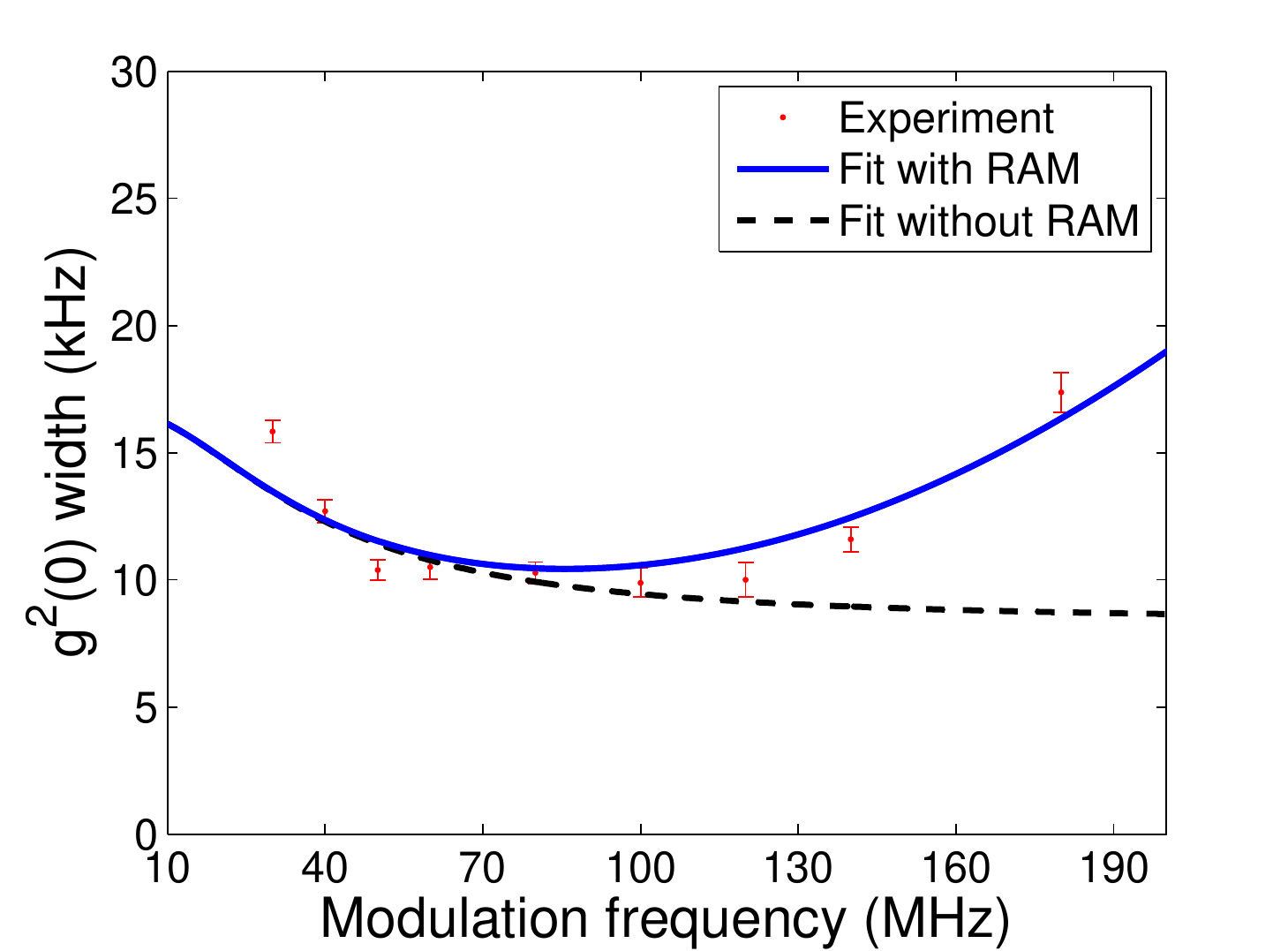}
\caption{(Color online) Measured $g^{(2)}(0)$ FWHM v.s. modulation frequency in the nonadiabatic regime. Laser power was 800 $\mu$W, and the transit width is $\gamma_2/\pi=63$ kHz.} \label{frequency.fig}
\end{figure}

Finally, we measured the $g^{(2)}(0)$ width in the nonadiabatic regime by current modulating a distributed feedback (DFB) Laser with about 2 MHz linewdith. Sub-lifetime-limited linewidth was also observed (see Fig.~\ref{frequency.fig}). The dependence of the $g^{(2)}(0)$ linewidth on input laser power as well as modulation depth (not shown) follows the same trend as in the adiabatic case. For a small modulation depth $\lambda=0.2$ and modulation frequency from 30 MHz to 180 MHz, the $g^{(2)}(0)$ width has a small variation in consistency with Eq.~(\ref{nonadiabatic}). Width increase at higher frequency is due to a combination of increasing influence of RAM (for signal is weaker at higher frequency according to Eq.~(\ref{nonadiabatic})) and slightly larger RAM as we found in the laser at higher modulation frequency. Calculation taking into account these facts and an overall scaling factor on the linewidth to include broadening from optical depth and other noises fits the data well. Ability to work at both low and high FM frequency is desirable for practical applications. For example, in CPT clock applications, the laser FM frequency can be chosen to be much higher than RF modulation, which makes the two modulations compatible.

The smallest attainable $g^{(2)}(0)$ width in practice is limited by noises. When modulation depth is so small that converted AM signal becomes comparable to noises such as laser intensity noise, RAM, electronics noise and eventually laser shot noise, the width starts to increase. In our transit-broadened system, we have achieved a minimal linewidth of 1.9 kHz with a signal-to-noise ratio (SNR)~\cite{SNR} of 62 for an average time of 16 ms; when optimized for the ratio of linewidth over SNR, CPT has a linewidth of 166 kHz  with SNR of 1500. The factor of 3.6 improvement in resolving power is mainly due to the distinct nature of the $g^{(2)}(0)$ observable. As well recognized, laser frequency noise induces amplitude noise and is a major noise source in precision spectroscopy including CPT-based applications~\cite{CPTreview,kitching}. However, the $g^{(2)}(0)$ measurement is much less sensitive to noise since only the noise locked to the laser modulation frequency is detected. An enlarged advantage of this method over traditional ones on resolving power is expected with improved FM techniques~\cite{FMspectroscopy}. In general, narrower resonance is at the price of a lower SNR, but linewidth considerably narrower than traditional ones is desirable because unknown systematics in resonance lineshapes make high ratio line splitting formidable~\cite{Phillips}. Therefore, large line narrowing factor combined with resolving power enhancement is the merit of this technique. In addition, we have numerically examined the possibility to sharpen multi-peaks, and find when CPT fails to completely separate two closely spaced peaks at its resolution limit, this method can clearly resolve them.

Creating quantum correlations between two bright optical fields and also among atomic spins using CPT is currently of great interest ~\cite{CPTsqueezing1,CPTsqueezing2}. Our work paves the way for the realization of such proposals, since inevitable laser frequency noise induces classical correlations in the same way as studied here~\cite{randomnoise}, and understanding its behavior is necessary for achieving quantum correlations. Furthermore, the sharp transition from correlation to anticorrelation observed here has a close resemblance to the behavior of quantum correlations predicted in ~\cite{HXM}, and are both resulted from the coherence induced sharp dispersion.

In conclusion, we have demonstrated a conceptually new way to obtain resonance linewidth far below the lifetime limit. Our methodology
might also be suitable for other level structures provided that an observable is identified which is only sensitive to phase information near a resonance center. Narrower linewidth combined with improved frequency sensitivity is useful for spectroscopy, precision measurements such as frequency standards~\cite{EITclock1,EITclock2}, magnetometry~\cite{kitching} and sensing based on resonance center location. Furthermore, the ability to sharpen  multi-peaks enables its application in resonance-based imaging~\cite{imaging}, where resonance with narrow width and high contrast leads to better resolution. These applications should be implementable in a wide variety of physical systems such as atoms~\cite{EITclock1}, molecules~\cite{molecules}, quantum dots~\cite{quantum-dots}, diamonds~\cite{Diamond}, and doped crystals \cite{rareearth1,rareearth2} etc., where Raman and CPT process exist.

We thank V. Vuleti\'{c}, E. Mikhailov, S. Wu, X. Hu, S. Du, M. Hohensee, S.Y. Zhu, W. Liu and M. Zhang for stimulating discussions. L.F. and P.L. contributed equally. This work is funded by the NBRPC (973 Program Grants No.s 2011CB921604 and 2012CB921604), NNSFC (Grants No. 10904018) and Fudan Univ.(Y.X.), by the Sherman Fairchild Foundation and the NBRPC (973 program Grants No.s 2011CBA00300 and 2011CBA00301) (L.J.), and by the AI-TF New Faculty Grant and the NSERC Discovery Grant (J.W.).


\begin{thebibliography}{99}
\bibitem{spectroscopy}
W.\ Demtr\"{o}der, \emph{Laser Spectroscopy} (Springer-Verlag Berlin Heidelberg New York, 1998).

\bibitem{Gawlik}
W.\ Gawlik, J.\ Kowalski, F.\ Tr\"{a}ger, and M.\ Vollmer,
Phys.\ Rev.\ Lett. \textbf{48}, 871 (1982).

\bibitem{Lam}
J.\ F.\ Lam, D.\ G.\ Steel, and R.\ A.\ McFarlane, Phys.\ Rev.\ Lett. \textbf{56}, 1679 (1986).

\bibitem{Gauthier}
D.\ Gauthier, Y.\ Zhu, and T.\ W.\ Mossberg, Phys.\ Rev.\ Lett. \textbf{66}, 2460 (1991).

\bibitem{Phillips}
H.\ Metcalf, W.\ Phillips, Opt.\ Lett. \textbf{5}, 540 (1980), and references therein.

\bibitem{Knight}
P.\ L.\ Knight and P.\ E.\ Coleman, J.\ Phys.\ B: Atom. Molec. Phys. \textbf{13}, 4345 (1980), and references therein.

\bibitem{Shimizu}
F.\ Shimizu, K.\ Umezu, H.\ Takuma, Phys.\ Rev.\ Lett. \textbf{47}, 825 (1981).

\bibitem{Scully}
H.\ Lee, P.\ Meystre, M.\ O.\ Scully, Phys.\ Rev.\ A \textbf{24}, 1914 (1981).

\bibitem{Albrecht}
M.\ A.\ Dugan, A.\ C.\ Albrecht, Phys.\ Rev.\ A \textbf{43}, 3877 (1991); ibid., 3922 (1991).

\bibitem{EITreview}
M.\ Fleischhauer, A. Imamoglu, J. P. Marangos, Rev.\ Mod.\ Phys.\
   \textbf{77}, 633 (2005).

\bibitem{CPTreview}
J.\ Vanier, Appl.\ Phys.\ B \textbf{81}, 421 (2005).

\bibitem{Camparo}
J.\ C.\ Camparo, J. G. Coffer, Phys.\ Rev.\ A \textbf{59}, 728
(1999).

\bibitem{CorrelationStudy}
L.\ S.\ Cruz, D. Felinto, J. G. Aguirre G¡äomez, M. Martinelli, P. Valente, A. Lezama, and P. Nussenzveig, Europ.\ Phys.\ J.\ D \textbf{41},
531 (2007).

\bibitem{Sautenkov}
V. A. Sautenkov, Y. V. Rostovtsev, and M. O. Scully, Phys.\ Rev.\ A \textbf{72}, 065801 (2005).

\bibitem{XiaoPRA}
Y. Xiao, T. Wang, M. Baryakhtar, M. Van Camp, M. Crescimanno, M. Hohensee, L. Jiang, D. F. Phillips,
M. D. Lukin, S. F. Yelin, and R. L. Walsworth, Phys.\ Rev.\ A \textbf{80}, 041805(R)(2009).

\bibitem{SNR}
For relevance in calculating resolving power, SNR is defined as the amplitude of the resonance devided by noise
measured at detuning equal to the half width of the resonance at half maximum.

\bibitem{kitching}
P. D. D. Schwindt, S. Knappe, V. Shah, L. Hollberg, J. Kitching, L.-A. Liew, J. Moreland, App.\ Phys.\ Lett. \textbf{85}, 6409 (2004).

\bibitem{FMspectroscopy}
M.\ Gehrtz, G.\ C.\ Bjorklund, E.\ Whittaker, J.\ Opt.\ Soc.\ Am.\ B \textbf{2}, 1510 (1985).

\bibitem{CPTsqueezing1}
A.\ Dantan, J.\ Cviklinski, E.\ Giacobino, M.\ Pinard, Phys.\ Rev.\ Lett. \textbf{97}, 023605 (2006).

\bibitem{CPTsqueezing2}
A.\ Sinatra, Phys.\ Rev.\ Lett. \textbf{97}, 253601 (2006).

\bibitem{randomnoise}
L.\ Feng, P.\ Li, T.\ Wang, Y.\ Xiao, to be published.

\bibitem{HXM}
F.\ Wang, X.\ Hu, W.\ Shi, Y.\ Zhu, Phys.\ Rev.\ A \textbf{81}, 033836 (2010).

\bibitem{EITclock1}
R.\ Santra, E.\ Arimondo, T.\ Ido, C.\ H.\ Greene, J.\ Ye, Phys.\ Rev.\ Lett. \textbf{94}, 173002 (2005).

\bibitem{EITclock2}
T. Zanon, S. Guerandel, E. de Clercq, D. Holleville, N. Dimarcq,
and A. Clairon, Phys.\ Rev.\ Lett. \textbf{94}, 193002 (2005).

\bibitem{imaging}
D. D. Yavuz and N. A. Proite, Phys.\ Rev.\ A \textbf{76}, 041802(R)(2007).

\bibitem{molecules}
A.\ Lazoudis, T.\ Kirova, E.\ H.\ Ahmed, L.\ Li, J.\ Qi, A.\ M.\ Lyyra, Phys.\ Rev.\ A \textbf{82}, 023812 (2010).

\bibitem{quantum-dots}
X.\ Xu, B.\ Sun, P.\ R.\ Berman, D.\ G.\ Steel, A.\ S.\ Bracker, D.\ Gammon, L.\ J.\ Sham, Nat.\ Phys. \textbf{4}, 692 (2008)

\bibitem{Diamond}
C. Santori, P. Tamarat, P. Neumann, J. Wrachtrup, D. Fattal, R. G. Beausoleil,
J. Rabeau, P. Olivero, A. D. Greentree, S. Prawer, F. Jelezko, and P. Hemmer,
Phys.\ Rev.\ Lett. \textbf{97}, 247401 (2006).

\bibitem{rareearth1}
E.\ Baldit, K.\ Bencheikh, P.\ Monnier, S.\ Briaudeau, J.\ A.\ Levenson, V.\ Crozatier, I.\ Lorgere, F.\ Bretenaker, J. L. Le Gouet,
O. Guillot-Noel, Ph. Goldner, Phys.\ Rev.\ B \textbf{81}, 144303 (2010).

\bibitem{rareearth2}
B.\ S.\ Ham, P.\ R.\ Hemmer, M.\ S.\ Shahriar, Opt.\ Commun. \textbf{144}, 227 (1997).

\end{thebibliography}
\end{document}